\documentclass[12pt]{article}
\usepackage{amssymb,amsfonts}
\usepackage{epsf,epsfig}
\begin{document}
\baselineskip 18pt

\title{One-magnon states and electron spin resonance in spin-ladders with singlet-rung ground state in a staggered magnetic
field}
\author{P.~N.~Bibikov}
\date{\it V.~A.~Fock Institute of Physics\\ Sankt-Petersburg State
University, Russia}

\maketitle

\vskip5mm

\begin{abstract}
One-magnon problem for spin-ladders with exact singlet-rung ground
state in arbitrary oriented constant and staggered magnetic fields
is studied. For two special configurations when the both fields
are parallel or perpendicular to each other the exact formulas for
eigenstates and eigenvectors are presented. In the parallele case
the one-magnon contribution to the ESR absorption line shape is
obtained. The latter is non-Lorentzian, has finite range and a gap
around resonance. For low fields both the range of the spectrum
and the width of the gap are controlled by the width of the
one-magnon zone in zero magnetic field. Some other interactions
usually contributing to the ESR line shape are also briefly
discussed.
\end{abstract}

\section{Introduction}

Effects of staggered magnetic field (SMF) in 1-D and 2-D spin
systems attract now a considerable interest \cite{1}. A violation
of the rotational symmetry induced by SMF mixes the states with
different polarizations while a violation of the translation
symmetry mixes the states with wave numbers connected by the
relation $k_2=k_1+\pi$.

In the present paper we study these effects basing on the special
spin-ladder model with exact singlet-rung ground state first
suggested in \cite{2}. An addition of rather small constant and
staggered magnetic fields do not destroy the vacuum however the
one-magnon problem transforms into the spectral problem for a
special $6\times6$ matrix. When the both fields are parallel the
remaining rotational invariance reduces the problem to three
spectral problems for $2\times2$ matrices. As it will be shown in
the Sect. 3 for the perpendicular configuration the problem also
simplifies and reduces to two $3\times3$ problems.

Violation of translation symmetry in spin-ladders was discussed in
\cite{3} in connection with anomaly in Raman scattering in the
ladder/chain system ${\rm Sr}_{14}{\rm Cu}_{24}{\rm O}_{41}$. It
was suggested that interaction constants of spin-ladder
Hamiltonian are modulated by the ladder-chain interaction.
Staggered magnetic field may be considered as a simple example of
such modulation.

Electron spin resonance (ESR) is a traditional experimental test
for small symmetry violations in spin systems \cite{4},\cite{5}.
In the present model the ESR transitions do not change a number of
magnons inducing only a change of polarization and a shift on
$\pi$ of a wave number.

As it will be shown in Sect. 3 in the presence of SMF magnons with
different polarizations have different dispersions. Owing to this
fact the ESR line shape broadens. For parallele constant and
staggered fields the exact formula for the one-magnon contribution
to line shape is calculated in Sect. 4. The obtained line shape
has a non-Lorentzian form, finite range and a gap around the
resonance. In Sect. 5 we discuss briefly effects of some other
symmetry violating terms such as dipole-dipole and
Dzyaloshinskii-Morya interactions.

Spin gap systems such as spin-ladders and Haldane spin-chains have
a significant feature. Their low-temperature properties with good
precision may be obtained from analysis of a few number of low
energy excitations \cite{6},\cite{7},\cite{8}. Since the
corresponding theoretical predictions may be tested by experiment,
the spin-gap systems probably give an ideal possibility for
theorists and experimenters to work in close connection. However
in contrast to Haldane chains the ESR in spin-ladders was not
studied so intensively. (See a very brief report in \cite{9}
devoted to ESR in ${\rm CaV}_2{\rm O}_5$). We believe that the
present paper will stimulate an interest to ESR in spin-ladder
materials and partially meet a lack in theoretical study of this
subject mentioned in \cite{4}.

\section{Spin Hamiltonian and magnon number operator}

General spin-ladder Hamiltonian has the following form:
\begin{equation}
{\cal H}=\sum_{n=-\infty}^{\infty}H^{(2)}_{n,n+1}+H^{(1)}_n,
\end{equation}
where
\begin{equation}
H_{n,n+1}^{(2)}=H^{leg}_{n,n+1}+H^{frust}_{n,n+1}+H^{cyc}_{n,n+1},\quad
H_n^{(1)}=H^{rung}_n+H^{Z}_n,
\end{equation}
and
\begin{eqnarray}
H^{leg}_{n,n+1}&=&J_{\|}({\bf S}_{1,n}\cdot{\bf S}_{1,n+1}+{\bf
S}_{2,n}\cdot{\bf
S}_{2,n+1}),\nonumber\\
H^{frust}_{n,n+1}&=&J_{frust}({\bf S}_{1,n}\cdot{\bf
S}_{2,n+1}+{\bf S}_{2,n}\cdot{\bf
S}_{1,n+1}),\nonumber\\
H^{cyc}_{n,n+1}&=&J_c(({\bf S}_{1,n}\cdot{\bf S}_{1,n+1})({\bf
S}_{2,n}\cdot{\bf S}_{2,n+1})+({\bf S}_{1,n}\cdot{\bf
S}_{2,n})({\bf S}_{1,n+1}\cdot{\bf S}_{2,n+1})\nonumber\\
&-&({\bf S}_{1,n}\cdot{\bf S}_{2,n+1})({\bf S}_{2,n}\cdot{\bf
S}_{1,n+1})),\nonumber\\
H^{rung}_n&=&J_{\bot}{\bf S}_{1,n}\cdot{\bf S}_{2,n},\nonumber\\
H^{Z}_n&=&-g\mu_B({\bf h}+(-1)^n{\bf h}_{st})\cdot({\bf
S}_{1,n}^z+{\bf S}_{2,n}^z).
\end{eqnarray}

Here ${\bf S}_{i,n}$ ($i=1,2;\,n=-\infty...\infty$) are
spin-$\frac{1}{2}$ operators associated with cites of the ladder,
{\bf h} and ${\bf h}_{st}$ are the constant and staggered magnetic
fields. Except the Sect. 5 we shall suppose that {\bf h} is
oriented along the $z$-axis.

In this paper we shall restrict the Hamiltonian (1)-(3) by two
additional conditions suggested in \cite{2}. The first one is the
following:
\begin{equation}
{[}{\cal H},{\cal Q}{]}=0,
\end{equation}
where ${\cal Q}=\sum_{n}\frac{3}{4}I+{\bf S}_{1,n}\cdot{\bf
S}_{2,n}$ is the magnon number operator or the sum of triplet-rung
projectors.

Let us denote by $|0\rangle_n$ and $|1\rangle_n^j$ ($j=-1,0,1$)
the singlet and triplet states associated with $n$-th rung. Then
the vector
\begin{equation}
|0\rangle=\prod_n|0\rangle_n,
\end{equation}
is the non degenerate zero-eigenstate of ${\cal Q}$. According to
(4) it is also an eigenstate of ${\cal H}$. Our second condition
demands the state (5) to be the ground state of ${\cal H}$.

As it was first shown in \cite{2} the condition (4) will be
satisfied if $J_{frust}=J_{\|}-\frac{1}{2}J_c$. The following
inequalities $J_{\bot}>2J_{\|}$, $J_{\bot}>\frac{5}{2}J_c$,
$J_{\bot}+J_{||}>\frac{3}{4}J_c$ also guarantee that for rather
small $h$ and $h_{st}$ the ground state has the form (5).

\section{One-magnon states in staggered magnetic field}

In the presence of SMF we have to suggest the following general
form of a one-magnon state:
\begin{equation}
|1,k;\nu\rangle_{magn}=\sum_{n=-\infty}^{\infty}\sum_{j=-1,0,1}\xi_j(k,\nu)e^{2ikn}|1,2n\rangle^j+
\eta_j(k,\nu)e^{ik(2n+1)}|1,2n+1\rangle^j,
\end{equation}
where $-\frac{\pi}{2}\leq k\leq\frac{\pi}{2}$ and
$|1,n\rangle^j=\ldots |0\rangle_{n-1}
|1\rangle_n^j|0\rangle_{n+1}\ldots$. The parameter $\nu=-1,0,1$
enumerates magnon polarizations.

The ${\rm Shr}\ddot{\rm o}{\rm dinger}$ equation for the
amplitudes (6) is equivalent to the following spectral problem:
\begin{equation}
A(k)\left(\begin{array}{c}
\xi\\
\eta
\end{array}\right)=E(k,\nu)\left(\begin{array}{c}
\xi\\
\eta
\end{array}\right),
\end{equation}
for the $6\times6$ matrix
\begin{equation}
A(k)=\left(\begin{array}{cc}
J_{\bot}-\frac{3}{2}J_c-g\mu_B(h\tilde
{\bf S}_z+{\bf h}_{st}\cdot\tilde{\bf S})&J_c\cos k\\
J_c\cos k&J_{\bot}-\frac{3}{2}J_c-g\mu_B(h\tilde {\bf S}_z-
{\bf h}_{st}\cdot\tilde{\bf S})\\
\end{array}\right).
\end{equation}
Here the $3\times3$ matrices $\tilde{\bf S}$ represent the
standard triple of $S=1$ spin operators.

For ${\bf h}_{st}\|{\bf h}$ the problem (7) has the following
system of solutions:
\begin{eqnarray}
\left(
\begin{array}{c}
\xi_+(k,\nu)\\
\eta_+(k,\nu)
\end{array}\right)&=&
\left(\begin{array}{c}
J_c\cos k\\
\nu g\mu_Bh_{st}+\sqrt{J_c^2\cos^2 k+\nu^2g^2\mu_B^2h_{st}^2}
\end{array}\right)\otimes e^{\nu},\nonumber\\
\left(\begin{array}{c}
\xi_-(k,\nu)\\
\eta_-(k,\nu)
\end{array}\right)
&=& \left(\begin{array}{c} \nu g\mu_Bh_{st}+\sqrt{J_c^2\cos^2
k+\nu^2g^2\mu_B^2h_{st}^2}\\
-J_c\cos k
\end{array}\right)\otimes e^{\nu},
\end{eqnarray}
with
\begin{equation}
E^{||}_{\pm}(k,\nu)=J_{\bot}-\frac{3}{2}J_c-\nu g\mu_B
h\pm\sqrt{J_c^2\cos^2k+\nu^2g^2\mu_B^2h_{st}^2},
\end{equation}
where $e^{\nu}\in{\mathbb C}^3$ and $\tilde{\bf S}^3e^{\nu}=\nu
e^{\nu}$ for $\nu=-1,0,1$.

For $J_c>0$ in the limit $h_{st}\rightarrow0$ the solution
$\xi_+(k,\nu),\,\eta_+(k,\nu)$ turns into magnon with wave number
$k$ \cite{2} while the solution $\xi_-(k,\nu),\,\eta_-(k,\nu)$
turns into magnon with wave number $k+\pi$. For $J_c<0$ these
limits change.

For ${\bf h}_{st}\bot{\bf h}$ the action of the matrix $A(k)$
decomposes the space ${\mathbb C}^6$ on two invariant subspaces
$V^{\bot}_+$ and $V^{\bot}_-$ generated by
\begin{equation}
v^{\nu}_{(\pm)}=\left(\begin{array}{c}
e^{\nu}\\
\mp(-1)^{\nu}e^{\nu}
\end{array}\right),\qquad \nu=-1,0,1,
\end{equation}
and the problem (7) splits on two problems for the matrices
\begin{equation}
A^{\bot}_{\pm}(k)=(J_{\bot}-\frac{3}{2}J_c\pm\frac{1}{3}J_c\cos
k)I-g\mu_Bh\tilde{\bf S}_z\pm 2J_c\cos k(\tilde{\bf
S}_z^2-\frac{2}{3}I)-g\mu_Bh_{st}\tilde {\bf S}_x,
\end{equation}
where $I$ is a unit matrix. Since
$A^{\bot}_-(k)=A^{\bot}_+(k+\pi)$ we may study only the "+"
problem in the interval $-\pi<k\leq\pi$. After the following
substitution
\begin{equation}
E^{\bot}_+(k,\nu)=J_{\bot}-\frac{3}{2}J_c+\frac{1}{3}J_c\cos
k+\frac{2}{3}J_c\varepsilon(k,\nu),
\end{equation}
the characteristic equation for the matrix $A^{\bot}_+(k)$
transforms to the form
\begin{equation}
\varepsilon^3(k,\nu)-3(\cos^2
k+\frac{1}{3}\gamma^2(h^2+h_{st}^2))\varepsilon(k,\nu)+2\cos
k(\cos^2k+\frac{1}{2}\gamma^2(h_{st}^2-2h^2))=0,
\end{equation}
where $\gamma=\frac{3g\mu_B}{2J_c}$. Now the standard
substitution: $\varepsilon(k,\nu)=\sqrt{\cos^2
k+\frac{1}{3}\gamma(h^2+h_{st}^2)}\lambda(k,\nu)$, reduces this
equation to the form $\lambda^3(k,\nu)-3\lambda(k,\nu)+2a(k)=0$
where
\begin{equation}
a(k)=\frac{(\cos^2 k+\frac{1}{2}\gamma^2(h_{st}^2-2h^2))\cos
k}{\sqrt{(\cos^2 k+\frac{1}{3}\gamma^2(h^2+h_{st}^2))^3}}.
\end{equation}
It may be easily proved that $|a(k)|\leq1$ and so putting
$a(k)=\sin3\alpha(k)$, where
$-\frac{\pi}{6}\leq\alpha(k)\leq\frac{\pi}{6}$ we obtain the
following representation for solutions of the Eq. (14)
\begin{equation}
\varepsilon(k,\nu)=2\sqrt{\cos^2
k+\frac{1}{3}\gamma(h^2+h_{st}^2)}\sin(\frac{1}{3}\arcsin{a(k)}+\frac{2\pi}{3}\nu),\quad
\nu=-1,0,1.
\end{equation}

The corresponding eigenvectors of the matrix $A^{\bot}_+(k)$ may
be represented as follows:
\begin{equation}
v^{\bot}(\varepsilon(k,\nu))=\left(\begin{array}{c}
\frac{\sqrt{2}}{2}\gamma h_{st}(\varepsilon(k,\nu)-\cos k-\gamma h)\\
\gamma^2 h^2-(\varepsilon(k,\nu)-\cos k)^2\\
\frac{\sqrt{2}}{2}\gamma h_{st}(\varepsilon(k,\nu)-\cos k+\gamma
h)
\end{array}\right).
\end{equation}

\section{One-magnon ESR line shape for ${\bf h}_{st}\|{\bf h}$}

According to \cite{10} the ESR line shape depends on the imaginary
part of magnetic susceptibility:
\begin{equation}
\chi''(\omega,T)\propto\sum_{E_a>E_b}{\rm
e}^{-\frac{E_a+E_b}{2kT}}\sinh\frac{\omega}{2kT}|\langle a|{\bf
S}^{tot}_x|b\rangle|^2\delta(E_a-E_b-\omega),
\end{equation}
where ${\bf S}^{tot}=\sum_n{\bf S}_{1,n}+{\bf S}_{2,n}$ is the
total spin of the ladder. In this section we shall obtain
$\chi_1''(\omega)$ the one-magnon contribution to the formula
(18). Calculation of $\chi_1''(\omega)$ for generally oriented
$\bf h$ and ${\bf h}_{\bf st}$ is very difficult even in the case
${\bf h}\bot{\bf h}_{\bf st}$. However for ${\bf h}_{\bf st}\|{\bf
h}$ it may be easily developed by straightforward substitution of
the formulas (9) and (10) into (18) (Of course with correct
normalization account of the vectors (9)). The delta function
removes summation in (18) and the main difficulty is to express
correctly $\cos k$ from $\omega$.

The calculations show that all transitions are grouped in pairs.
Contribution of each pair lies in the corresponding frequency
interval and have one of the two possible forms,
\begin{equation}
\chi_{\pm}''(\omega,T)=\frac{\Gamma^2\cosh{\frac{1}{2kT}(\frac{\Gamma^2}{4(\omega\mp
\omega_{res})}\pm \omega_{res})}}{\frac{\Gamma^2}{4}+(\omega\pm
\omega_{res})^2}\sinh{\frac{\omega}{2kT}},
\end{equation}
where $\omega_{res}=g\mu_Bh$ and $\frac{\Gamma}{2}=g\mu_Bh_{st}$.

The two pairs $E^{||}_+(k,1)\rightarrow E^{||}_-(k,0)$,
$E^{||}_+(k,0)\rightarrow E^{||}_-(k,-1)$ and
$E^{||}_-(k,0)\rightarrow E^{||}_-(k,-1)$,
$E^{||}_+(k,1)\rightarrow E^{||}_+(k,0)$ contribute
$\chi_-(\omega)$ in the frequency interval
$\omega_-\leq\omega\leq\omega_{res}-\delta\omega$. The two pairs
$E^{||}_-(k,1)\rightarrow E^{||}_-(k,0)$,
$E^{||}_+(k,0)\rightarrow E^{||}_+(k,-1)$ and
$E^{||}_-(k,1)\rightarrow E^{||}_+(k,0)$,
$E^{||}_-(k,0)\rightarrow E^{||}_+(k,-1)$ also contribute
$\chi_-(\omega)$ but in the frequency interval
$\omega_{res}+\delta\omega\leq\omega\leq\omega_+$. However the two
pairs $E^{||}_-(k,-1)\rightarrow E^{||}_-(k,0)$,
$E^{||}_+(k,0)\rightarrow E^{||}_+(k,1)$ and
$E^{||}_-(k,-1)\rightarrow E^{||}_+(k,0)$,
$E^{||}_-(k,0)\rightarrow E^{||}_+(k,1)$ contribute
$\chi_+''(\omega)$ in the interval
$\omega_-\leq\omega\leq\omega_{res}-\delta\omega$. Here
\begin{eqnarray}
\delta\omega&=&\sqrt{J_c^2+g^2\mu_B^2h_{st}^2}-|J_c|,\nonumber\\
\quad\omega_{\pm}&=&g\mu_Bh\pm|J_c|\pm\sqrt{J_c^2+g^2\mu_B^2h_{st}^2}.
\end{eqnarray}

Finitely we may write the following formula
\begin{eqnarray}
\chi_1''(\omega,T)&\propto&
\chi_-(\omega,T)+\chi_+(\omega,T),\qquad \omega_{min}\leq\omega\leq\omega_{res}-\delta\omega;\nonumber\\
\chi_1''(\omega,T)&\propto&\chi_-(\omega,T),\qquad
\omega_{res}+\delta\omega \leq\omega\leq\omega_{max},
\end{eqnarray}
where $\omega_{min}={\rm max}(0,\omega_-)$ and
$\omega_{max}=\omega_+$.

We see that the obtained line shape may be interpreted as twin
asymmetric peaks separated by the gap. However it seems to us more
fine to interpret it as a single-resonance contour with the gap
around the resonance. The range of the spectrum and the width of
the gap are controlled by {\it cutoff} parameters
$\omega_{max}-\omega_{min}$ and $2\delta\omega$.

For low magnetic fields $h,h_{st}\ll\frac{|J_c|}{g\mu_B}$ a good
approximation will be $\omega_{min}=0$, $\omega_{max}=2|J_c|$ and
$\delta\omega=\frac{g^2\mu_B^2h_{st}^2}{2|J_c|}$. In this case
according to (10) $2|J_c|=\Delta E_{magn}$ where the latter is the
width of the one-magnon zone. The parameter $\Delta E_{magn}$ does
not appear in the formula (19) but according to the following
approximate formulas
\begin{equation}
\delta\omega\simeq\frac{g^2\mu_B^2h_{st}^2}{\Delta
E_{magn}},\qquad \omega_{max}\simeq\Delta E_{magn}.
\end{equation}
it has the sense of a {\it scale parameter}. In contrast to
"mystical" ones often introduced ad hoc in Quantum Field Theory
this parameter appears {\it dynamically} and has the clear
physical interpretation and manifests the high-energy (spatial
magnon dynamic) level at the low-energy (polarization magnon
dynamic) one.

For $\omega_{res}\ll\omega\ll kT$ the magnetic susceptibility (21)
has the following asymptotic:
\begin{equation}
\chi_1''(\omega,T)\propto\frac{1}{kT\omega}.
\end{equation}
For $\omega\rightarrow\omega_{res}\pm\delta\omega$ asymptotics are
the following:
\begin{equation}
\chi_1''(\omega_{res}\pm(\delta\omega+\epsilon),T)\propto\frac{1}{kT}{\rm
e}^{-(\frac{\Delta
E_{magn}}{g\mu_Bh_{st}})^2\frac{\epsilon}{2kT}}.
\end{equation}

The similar asymptotics at
$\omega\rightarrow\omega_{res}\pm\delta\omega$ has the Gauss
function: \\
$\chi_{Gauss}''(\omega_{res}\pm(\delta\omega+\epsilon))\propto\frac{1}{kT}{\rm
e}^{-(\frac{\Delta
E_{magn}}{g\mu_Bh_{st}})^4\frac{(\delta\omega+\epsilon)^2}{4kT\Delta
E_{magn}}}$.

\section{Dipole-dipole and Dzyaloshinskii-Moria interactions}
Traditionally ESR is applied for measuring dipole-dipole and
Dzyaloshinskii-Moria interactions. In this section we shall
briefly discuss both of them restricting ourselves on the case
when the condition (4) is satisfied.

Dipole-dipole interaction along rungs induce an appearance of the
following term
\begin{equation}
H^{rung-dip}_n=D_{\bot} ({\bf S}^z_{1,n}{\bf
S}^z_{2,n}+\frac{1}{4}I)=\frac{D_{\bot}}{2}({\bf S}_{1,n}^z+{\bf
S}_{2,n}^z)^2,\quad D_{\bot}>0.
\end{equation}
In the absence of SMF the Hamiltonian (1)-(3), (25) is
translational invariant and a one-magnon state has the form:
\begin{equation}
|1,k;\nu\rangle_{magn}=\sum_{n=-\infty}^{\infty}\sum_{j=-1,0,1}\zeta_j(k,\nu)e^{ikn}|1,n\rangle^j,
\end{equation}
where the amplitudes $\zeta_j(k,\nu)$ satisfy the following ${\rm
Shr}\ddot{\rm o}{\rm dinger}$ equation:
\begin{equation}
(J(k)-g\mu_B{\bf h}\tilde{\bf S}+\frac{D_{\bot}}{2}\tilde{\bf
S}_z^2)\zeta(k,\nu)=E(k,\nu)\zeta(k,\nu).
\end{equation}
Here $J(k)=J_{\bot}-\frac{3}{2}J_c+J_c\cos k$. The cubic equation
corresponding to the spectral problem (27) may be solved by the
same method as the Eq. (14). Effect of the rung-dipole interaction
reveals in splitting of resonance lines without any broadening.

The following terms
\begin{eqnarray}
H_{n,n+1}^{leg-dip}&=&D_{\|}({\bf S}^{z}_{1,n}+{\bf S}^{z}_{2,n})({\bf S}^{z}_{1,n+1}+{\bf S}^{z}_{2,n+1 }),\\
H_{n,n+1}^{DM}&=&{\bf J}_{DM}\cdot({[}{\bf S}_{1,n}\times{\bf
S}_{1,n+1}{]}+{[}{\bf S}_{2,n}\times{\bf S}_{2,n+1}{]}+{[}{\bf
S}_{1,n}\times{\bf S}_{2,n+1}{]}\nonumber\\
&+&{[}{\bf S}_{2,n}\times {\bf S}_{1,n+1}{]}),
\end{eqnarray}
preserve the condition (4) and in some sense may be interpreted as
dipole-dipole and Dzyaloshinskii-Morya interactions along legs and
diagonals. However these terms do not contribute in the one magnon
sector.

\section{Conclusions}
In the present paper we have obtained the exact formulas for
one-magnon excitations in spin-ladders with exact singlet-rung
ground state and staggered magnetic field oriented parallel or
perpendicular to the constant magnetic field. In the parallele
case we have calculated the one-magnon contribution to the ESR
line shape. The latter is non-Lorentzian. It has the finite range
and the gap around the resonance. Near the gap the line shape is
Gaussian. For low magnetic fields both the width of the gap and
the range of the spectrum are controlled by the the scale of the
one-magnon zone. The presented results allow to conclude that the
effects of staggered magnetic field are not perceptible in the
spin-ladder material ${\rm CaV}_2{\rm O}_5$ \cite{9}.

\section{Acknowledgments}

The author is very grateful for the discussion with S.~A.~Paston.

\end{document}